\documentclass[aps,pra,twocolumn,preprintnumbers,amsmath,amssymb,footinbib]{revtex4}

\usepackage{graphicx}
\usepackage{dcolumn}
\usepackage{bm}

\usepackage[utf8]{inputenc}
\usepackage[T1]{fontenc}
\usepackage{mathptmx}
\usepackage{xcolor}

\begin{document}

\title{Comparison of the Tetrahedron Method to Smearing Methods for the Electronic Density of States}

\author{Michael Y. Toriyama$^{1}$}
\author{Alex M. Ganose$^{2}$}
\author{Max Dylla$^{1}$}
\author{Shashwat Anand$^{1}$}
\author{Junsoo Park$^{2}$}
\author{Madison K. Brod$^{1}$}
\author{Jason Munro$^{3}$}
\author{Kristin A. Persson$^{4,5}$}
\author{Anubhav Jain$^{2}$}
\author{G. Jeffrey Snyder$^{1}$}
\affiliation{$^{1}$Department of Materials Science and Engineering, Northwestern University, Evanston, IL 60208, USA.}
\affiliation{$^{2}$Energy Technologies Area, Lawrence Berkeley National Laboratory, Berkeley, CA 94720, USA.}
\affiliation{$^{3}$The Materials Science Division, Lawrence Berkeley National Laboratory, Berkeley, CA 94720, USA.}
\affiliation{$^{4}$Department of Materials Science and Engineering, University of California, Berkeley, Berkeley, CA 94720, USA.}
\affiliation{$^{5}$The Molecular Foundry, Lawrence Berkeley National Laboratory, Berkeley, CA 94720, USA.}

\date{\today}

\begin{abstract}
The electronic density of states (DOS) highlights fundamental properties of materials that oftentimes dictate their properties, such as the band gap and Van Hove singularities. In this short note, we discuss how sharp features of the density of states can be obscured by smearing methods (such as the Gaussian and Fermi smearing methods) when calculating the DOS. While the common approach to reach a ``converged'' density of states of a material is to increase the discrete k-point mesh density, we show that the DOS calculated by smearing methods can appear to converge but not to the correct DOS. Employing the tetrahedron method for Brillouin zone integration resolves key features of the density of states far better than smearing methods.
\end{abstract}

\maketitle

Perhaps the simplest descriptor of the electronic structure of a material is the density of states, which condenses many fundamental electrical and optical properties into a single informative diagram. Features such as the band gap and the effective mass of carriers are directly related to transport properties of the material, whereas sharp peaks (i.e. Van Hove singularities) provide critical information about optical properties such as the dielectric constant.

Oftentimes, computational methods based on first-principles are used to calculate the density of states of materials. However, computational approaches require a trade-off between computational cost and accuracy, contextualizing the notion of ``convergence'' in which the property of interest is calculated with increasing computational cost until the property no longer appears to change within the desired accuracy. For the density of states, convergence is especially important for generating salient features of the electronic structure. Such numerical artefacts may arise from the technique used to calculate the density of states, which can broadly be divided into two categories: ``smearing methods'' and the ``tetrahedron method''.

Smearing methods (as described mathematically in the Appendix) fix a continuous function at each band and k-point to approximate the density of states. The tetrahedron method on the other hand divides the Brillouin zone into tetrahedra, calculates the eigenenergies at the corners of each tetrahedron, and linearly interpolates the eigenenergies inside of each tetrahedron to perform the integration. The tetrahedron method is reminiscent of the trapezoidal rule for approximating the integral of single-variable functions. As is the case with single-variable functions where a linear interpolation overestimates regions of positive curvature and underestimates regions of negative curvature, the linear-tetrahedron method is prone to similar errors. In response, a correction to the integration weights was introduced by Bl\"{o}chl,\cite{TetrahedronMethod} and the corrected Brillouin zone integration method is called the tetrahedron method with Bl\"{o}chl corrections.

\begin{figure}[!t]
    \centering
    \includegraphics[width=0.45\textwidth]{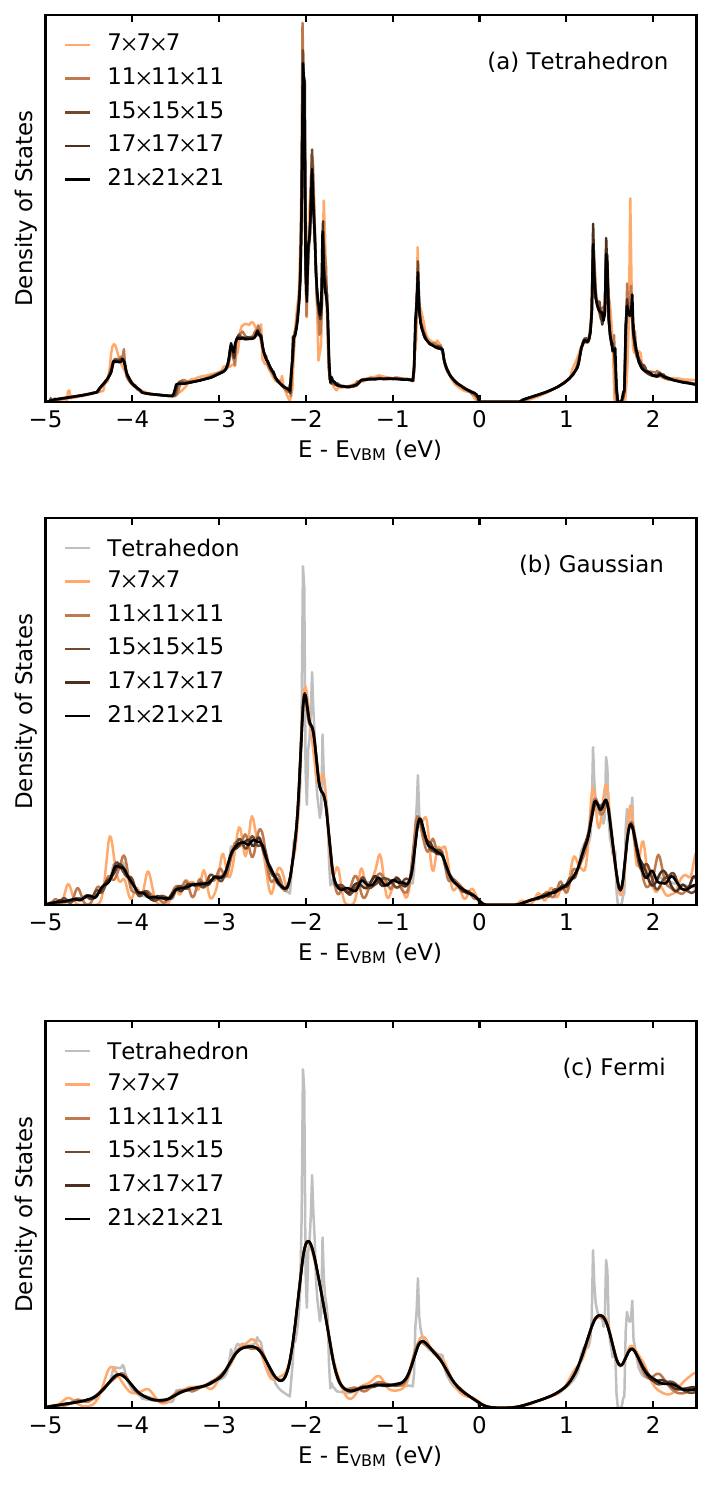}
    \caption{Density of states of TiNiSn using different k-point sampling methods. The width parameter is fixed to 50 meV for the smearing methods. The tetrahedron method (a) shows a clear band gap, van Hove singularities and flat DOS regions even with a 7×7×7 k-point mesh. The Gaussian (b) and Fermi (c) smearing methods appear to converge the density of states above a mesh size of 15, but even the converged density of states blurs the features seen in the tetrahedron method and can produce artificial features even at a much denser k-point mesh.}
    \label{Fig:MeshConvergence}
\end{figure}

In this short note, we compare the density of states that are calculated using smearing methods and the tetrahedron method. We use first-principles Density Functional Theory (DFT) to calculate the density of states of the half-Heusler compound TiNiSn (Materials Project ID: mp-924130).\cite{Jain2013} We compare the density of states calculated using the tetrahedron method with Bl\"{o}chl corrections\cite{TetrahedronMethod} against that calculated using smearing methods. DFT calculations were performed using the Vienna ab-initio simulation package (VASP).\cite{DFT_1, DFT_2, DFT_3} Exchange-correlation effects were treated using the Perdew-Burke-Ernzerhof functional\cite{1996_PBE} for all calculations. We use the recommended Ti\_pv, Ni\_pv, and Sn\_d pseudopotentials distributed by VASP. Each structure was relaxed using a 520 eV energy cutoff and a $\Gamma$-centered grid to sample k-points in the Brillouin zone.\cite{MonkhorstPack} Spin-orbit coupling effects were not included in the calculations. The density of states is calculated using 5001 energy bins. We employ the Gaussian and Fermi smearing methods, where we use a smearing width of 0.05 eV.

As shown in Figure \ref{Fig:MeshConvergence}, key analytical features representative of the electronic structure of a material are clearly evident when the tetrahedron method with Bl\"{o}chl corrections is employed. Sharp Van Hove peaks such as those at 0.8 eV and 2 eV below the valence band maximum are visible from the density of states calculated using the tetrahedron method, whereas the peaks are obscured by Gaussian and Fermi smearings. Gaps in the density of states, such as the one at 1.6 eV above the valence band maximum observed by the tetrahedron method, close when smearing methods are employed.

\begin{figure}
    \centering
    \includegraphics[width=0.45\textwidth]{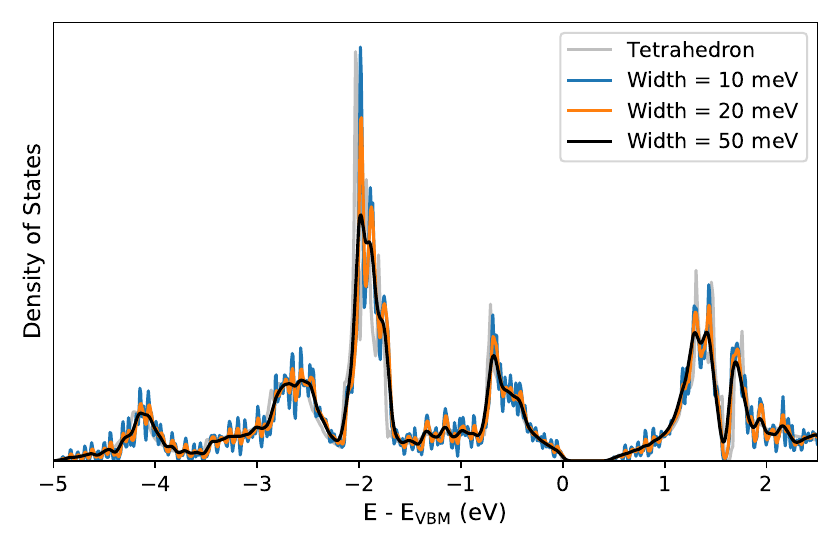}
    \caption{The density of states calculated using the Gaussian smearing method with different smearing parameters, using a dense k-point mesh of $21 \times 21 \times 21$.}
    \label{Fig:Gaussian_SmearingWidths}
\end{figure}

Smearing methods should in principle reach the true density of states by increasing the computational cost by e.g. increasing the k-point mesh density. However, as shown in Figure \ref{Fig:MeshConvergence}, stark features such as Van Hove singularities and the band edge shape are obscure even at higher k-point densities using smearing methods without carefully adjusting the smearing parameters, whereas they appear clearly with the tetrahedron method. Although the density of states of the Gaussian and Fermi smearing methods separately converge to a stable distribution as the k-point density increases, they do not approach the density of states calculated using the tetrahedron method. Moreover, the width parameter of the smearing methods must be carefully chosen to reach the density of states calculated using the tetrahedron method. As shown in Figure \ref{Fig:Gaussian_SmearingWidths}, using a comparatively large smearing width of 50 meV obscures the peaks at $\sim$ 2 eV below the valence band maximum, whereas using a small width of 10 meV introduces spurious noise. It is therefore clear from this comparison that the tetrahedron method is the preferred computational method for evaluating the electronic structure of a semiconductor, since many features of the electronic structure are markedly expressed.

We stress the importance of the Brillouin zone integration method over the k-point mesh density in this note. Specifically, the tetrahedron method for Brillouin zone integration is preferred over smearing methods to generate sharp qualities of the density of states. Although many in the community already follow this recommendation, they are important to re-emphasize as the user community of electronic structure methods grows.

\section*{Appendix}

The density of states $g(E)$ is calculated as
\begin{align} \label{TotalElectrons}
g(E) = \frac{1}{V} \sum_n \int_{V} \delta(E - \epsilon_{n,\bf{k}}) d\bf{k}
\end{align}
where $\epsilon_{n,\bf{k}}$ is the energy of band $n$ at k-point $\bf{k}$, and $V$ is the volume of the reciprocal primitive cell. The Gaussian and Fermi smearing methods approximate the $\delta(E - \epsilon_{n,\bf{k}})$ function in the following ways:

\noindent Gaussian smearing:
\begin{align}
\delta(E - \epsilon_{n,\bf{k}}) \approx \frac{1}{\sigma \sqrt{\pi}} e^{-\left( \frac{E - \epsilon_{n,\bf{k}}}{\sigma} \right)^2}
\end{align}

\noindent Fermi smearing:
\begin{align}
\delta(E - \epsilon_{n,\bf{k}}) \approx \frac{e^{-\frac{E - \epsilon_{n,\bf{k}}}{\sigma}} }{\sigma \left( 1 + e^{-\frac{E - \epsilon_{n,\bf{k}}}{\sigma}} \right)^2}
\end{align}

\section*{Acknowledgements}

We acknowledge support from NSF DMREF award \#1729487. MYT acknowledges support from the United States Department of Energy through the Computational Science Graduate Fellowship (DOE CSGF) under Grant Number DE-SC0020347. This research was supported in part through the computational resources and staff contributions provided for the Quest high performance computing facility at Northwestern University which is jointly supported by the Office of the Provost, the Office for Research, and Northwestern University Information Technology. Work by AG, JP, and AJ was supported by the U.S Department of Energy, Office of Basic Energy Sciences, and the Early Career Research Program. JM and KAP acknowledge support from the U.S. Department of Energy, Office of Science, Office of Basic Energy Sciences, Materials Sciences and Engineering Division under contract no. DE-AC02-05-CH11231 (Materials Project program KC23MP). This research used resources of the National Energy Research Scientific Computing Center (NERSC), a U.S. Department of Energy Office of Science User Facility operated under Contract No. DE-AC02-05CH11231.

\bibliography{bibliography}

\end{document}